\documentclass[journal=jctcce,manuscript=article,layout=twocolumn]{achemso}

\usepackage{amsmath} 
\usepackage[capitalize]{cleveref} 
\usepackage[labelfont=bf]{caption}
\setkeys{acs}{maxauthors=5, etalmode=truncate}
\usepackage{setspace}

\usepackage{textgreek}

\renewcommand{\vec}{\boldsymbol}

\let\oldmaketitle\maketitle
\let\maketitle\relax

\author{Michael J. Hartmann}
\author{Yuvraj Singh}
\affiliation{Department of Chemistry, New York University}
\author{Eric Vanden-Eijnden}
\affiliation{Courant Institute of Mathematical Sciences, New York University}
\email{eve2@cims.nyu.edu}
\author{Glen M. Hocky}
\affiliation{Department of Chemistry, New York University}
\email{hockyg@nyu.edu}

\title[FISST]{Infinite Switch Simulated Tempering in Force (FISST)}

\begin{document}

\twocolumn[
\begin{@twocolumnfalse}
\oldmaketitle
\begin{abstract}
\noindent
Many proteins in cells are capable of sensing and responding to piconewton scale forces, a regime in which conformational changes are small but significant for biological processes. 
In order to efficiently and effectively sample the response of these proteins to small forces, enhanced sampling techniques will be required.
In this work, we derive, implement, and evaluate an efficient method to simultaneously sample the result of applying any constant pulling force within a specified range to a molecular system of interest.
We start from Simulated Tempering in Force, whereby force is applied as a linear bias on a collective variable to the system's Hamiltonian, and the coefficient is taken as a continuous auxiliary degree of freedom.
We derive a formula for an average collective-variable-dependent force, which depends on a set of weights, learned on-the-fly throughout a simulation, that reflect the limit where force varies infinitely quickly.
These weights can then be used to retroactively compute averages of any observable at any force within the specified range. 
This technique is based on recent work deriving similar equations for Infinite Switch Simulated Tempering in Temperature, that showed the infinite switch limit is the most efficient for sampling. 
Here, we demonstrate that our method accurately and simultaneously samples molecular systems at all forces within a user defined force range, and show how it can serve as an enhanced sampling tool for cases where the pulling direction destabilizes states of low free-energy at zero-force.
This method is implemented in, and will be freely-distributed with, the PLUMED open-source sampling library, and hence can be readily applied to problems using a wide range of molecular dynamics software packages.
\end{abstract}
\end{@twocolumnfalse}
]

\section{Introduction}
Mechanical forces acting on the molecular scale play a crucial role across biology, from driving essential processes such as cell migration to determining the emergent macroscale properties of biological materials\cite{bao2003cell,davis2009force,roca2017quantifying, heisenberg2019disease, patapoutian2018gpr68}.
While the response of macroscopic systems to force can be measured by rheological techniques and often matched to theories of elasticity or viscous flow,\cite{ribbeck2017rheology, saintillan2018rheology} understanding the response of microscopic systems to force is more challenging.
Significant progress has been made through pioneering single-molecule force spectroscopy studies, which have given insight into the folding landscape of proteins, the kinetics of protein-protein interactions, and the behavior of molecular motors \cite{fernandez2004force, Vera2016ppis, Duwez2018foldamers}.

Molecular dynamics (MD) simulations are often capable of representing the equilibrium behavior of a system, and therefore are a key tool to elucidate the detailed, molecular-scale picture of what underlies important chemical and biological processes.\cite{allen2019md, hocky2016actin, tielman2019plis}
Schulten and others pioneered the use of Steered Molecular Dynamics (SMD) to predict the behavior of molecules in single molecule pulling experiments, where an external force is applied within a molecular dynamics simulation in a way that mimics common experimental setups.\cite{isralewitz2001steered,park2003free}.
Although these SMD simulations and the experiments they mimic are performed out-of-equilibrium, it is in principle possible to use non-equilibrium fluctuation theorems to extract equilibrium information from an ensemble of trajectories. \cite{hummer2001free,liphardt2002equilibrium,park2003free,zhang2011reconstructing}.
However, in order to observe empirically relevant structural changes (e.g. protein unfolding) within an achievable simulation time scale, SMD pulling must be performed with unphysically large forces applied to the system.\cite{sahoo2019ubiquitin}
These large forces then lead to poor agreement with the experiments they were designed to simulate and a dependence of the result on the pulling rate.\cite{daday2019smdrate}
To overcome these limitations, one can reduce the number of degrees of freedom and artificially smooth the free energy landscape by using coarse-grained models, which effectively decreases the timescale of the targeted process.\cite{Marrink2014coarsegraining}
Alternatively, one could use enhanced sampling simulations in conjunction with SMD at lower pulling forces/rates to more quickly sample a molecules conformations \cite{klimov1999stretching,mucksch2016accelerating}. 
These combined approaches are challenging because they require techniques for sampling non-equilibrium trajectories, and typically are more difficult to converge for large systems than standard equilibrium sampling methods \cite{dickson2011flow}. 

In this work, we focus on systems under a small constant mechanical load on the order of a few to tens of piconewtons (pN), a regime known to initiate and drive many important biological processes.\cite{grashoff2016tension}
Although applying a pulling force generally drives the system out of equilibrium, thermodynamically speaking, applying a small constant force simply creates a new, tilted, energy landscape on which the system will equilibrate \cite{bustamante2004mechanical}.
Moreover, small applied forces are expected to be near the linear-response regime, and would simply change the weight of the conformations observed at equilibrium rather than drive large conformational changes often studied by single molecule force probes and SMD simulations.
Because of these factors, standard equilibrium sampling methods such as Parallel Tempering should adequately probe the effect of these forces on the resulting conformational ensemble. 
Here, we investigate a method in which small applied forces can be used to simultaneously obtain equilibrium information about a molecular system while also accelerating sampling. 

Martinsson \textit{et al.} have recently developed a useful enhanced sampling method called Infinite Switch Simulated Tempering (ISST).\cite{martinsson2019simulated}
They show that the most efficient way of performing simulated tempering, where temperature is a dynamical variable in the simulation, occurs in the limit where the temperature can change infinitely quickly. 
In this regime, an effective configuration-dependent temperature is learned and used to propagate the dynamics.
Information about the system at any temperature in the chosen temperature range between $T_{\min}$ and $T_{\max}$ can be obtained {\em post facto} using weights calculated on-the-fly during the simulation. 

In this work, we derive the force-equivalent of the ISST method for a dynamic force variable, which we term FISST. 
FISST allows us to run simulations over a user defined range of forces, and by learning the `weights' for each force on-the-fly, quantitatively reconstruct the probability density function of a given observable at any force within the force range, effectively gaining information about the $N_f$ different forces one wants to study with a single simulation.
Importantly, because FISST is a collective-variable-based method, it only depends on an intensive quantity of the system, hence its effectiveness does not deteriorate with system size or dimensionality. 
We illustrate the performance of FISST for a number of test systems ranging in complexity, including a simple 2D analytical potential, a chain of beads with i/i+4 interactions that favors a degenerate left and right handed helical configurations at zero force, and deca-alanine in water.
In addition, we attempt to quantify the amount of information gained using FISST over traditional equilibrium sampling methods and comment on the prospect of FISST as an enhanced sampling method.
FISST is implemented as a module in the open-source PLUMED package,\cite{Bussi2014plumed2,bonomi2019promoting} a plug-in for many of the most popular simulation packages, and can therefore be immediately applied to virtually any system of interest.

\section{Theory and Methods}
\label{sec:theory}
\subsection{Simulations under constant force}
Assume that the system under zero force has the Hamiltonian:
\begin{equation}
    H(\vec{p},\vec{q}) = \tfrac12 \vec{p}^T M^{-1} \vec{p} + U(\vec{q}),
\end{equation}
where $\vec{q}$ and $\vec{p}$ represent the position and momenta of the particles in the system, $M$ is the mass matrix, and $U(\vec{q})$ the potential. 
Then the system's Hamiltonian with a force can be written as:
\begin{equation}
    H_F(\vec{p},\vec{q}) = \tfrac12 \vec{p}^T M^{-1} \vec{p} + U(\vec{q}) - F Q(\vec{q}).
\end{equation}
where 
$Q(\vec{q})$ is a collective variable (CV), defined here as a function of particle positions (although this could be generalized). 
In this equation, a positive $F$ corresponds to pulling (i.e. a larger $Q$ will be preferred for $F>0$). 
It is evident that as long as $F$ does not vary in time, then any standard constant-temperature MD, MC, or enhanced sampling method can be applied to sample configurations from the equilibrium Boltzmann distribution with density $\propto e^{-\beta H_F}$, where $\beta=(k_B T)^{-1}$, $k_B$ is Boltzmann's constant, and $T$ is the temperature of the system.

\subsection{Simulated Tempering in Force}
One way to do tempering in force is to do Hamiltonian Replica Exchange \cite{fukunishi2002hamiltonian}, with discrete forces applied to collective variables. 
To study $N_F$ different forces in the range $F_{\min}<F_i<F_{\max}$, we would simulate $N_F$ copies of our system with Hamiltonians given by: 
\begin{equation}
    H_i(\vec{p},\vec{q}) = \tfrac12 \vec{p}^T M^{-1} \vec{p} + U(\vec{q}) - F_i Q(\vec{q}).
\end{equation}
Monte Carlo exchanges between replicas are done periodically, with a Metropolis acceptance rate of $P_\mathrm{exchange}=\min\{1,\exp(-\beta(F_i-Fj)(Q_i-Qj))\}$.

Alternatively, one could perform the equivalent of a continuous version of simulated tempering \cite{marinari1992simulated,martinsson2019simulated}, in which case $F$ becomes a continuous extra degree of freedom. It is then possible to perform Langevin Dynamics (LD)  such that the following probability density for each configuration $(\vec{q},F)$ is sampled:
\begin{equation}
    \rho(\vec{q},F) = C^{-1}(\beta) \omega(F) e^{-\beta U(\vec{q}) + \beta F Q(\vec{q})}.
    \label{eqn:fulldensity}
\end{equation}
Here $\omega(F)$ is a weight function to be  specified (more on this below) which is positive for $F_{\min}<F<F_{\max}$ and zero outside that range (such that these forces are not accessible), and 
\begin{align}
    C(\beta) &=  \int_{F_{\min}}^{F_{\max}} dF\, \omega(F) \int d\vec{q} \, e^{-\beta U(\vec{q}) + \beta F Q(\vec{q})}
             \label{eqn:gaussint}\\
             & \equiv \int_{F_{\min}}^{F_{\max}} dF \, \omega(F) Z_q(F)
             \label{eqn:part}.
\end{align}
where we have defined the partition function
\begin{equation}
    Z_q(F) \equiv \int d\vec{q}\, e^{-\beta U(\vec{q}) + \beta F Q(\vec{q})}.
    \label{eqn:zq}
\end{equation}

\subsection{The Infinite Switch Limit}
As discussed above, the arguments of Ref.~\citenum{martinsson2019simulated} suggest that the most efficient sampling scheme occurs in the infinite switch limit, i.e. when the mass of the fictitious ``force-momentum'' becomes 0. 
In this limit, we can write an alternative LD scheme to sample a phase space density for $\vec{q}$ where force has been integrated over by the fast dynamics of $F$,
\begin{equation}
    \bar{\rho}(\vec{q}) = \frac{ \int_{F_{\min}}^{F_{\max}}dF\, \omega(F) e^{-\beta U(\vec{q}) + \beta F Q(\vec{q})}}{ \int_{F_{\min}}^{F_{\max}}dF\, Z_q(F) \omega(F)}
    \label{eqn:rhobar}
\end{equation}
This scheme is given by the following equations, 
\begin{align}
\frac{d\vec{q}}{d t} &= M^{-1}\vec{p} \label{eqn:LDq}\\
\frac{d\vec{p}}{d t} &= -\nabla U + \bar{F}(Q) \nabla Q - \gamma \vec{p} + \sqrt{2 \gamma \beta^{-1} }M^{1/2} \vec{\eta}
\label{eqn:LDp}
\end{align}
where $\vec{\eta}$ is a white-noise with corelation $\langle \eta_i(t)\eta_j(s)\rangle = \delta_{i,j} \delta(t-s)$, $\gamma$ is the friction coefficient, and the dynamical variable $F$ from the extended LD scheme has been replaced by the average $\bar{F}(Q)$ given by,
\begin{equation}
\begin{split}
    \bar{F}(Q) &= \int_{F_{\min}}^{F_{\max}}dF\, F \rho(F | \vec{q}) \\
               &\equiv\frac{ \int_{F_{\min}}^{F_{\max}}dF\, e^{-\beta U(\vec{q})+\beta F Q(\vec{q})} \omega(F) F }{\int_{F_{\min}}^{F_{\max}}dF\, e^{-\beta U(\vec{q})+\beta F Q(\vec{q})} \omega(F)}  \\
               &=\frac{ \int_{F_{\min}}^{F_{\max}} dF\,e^{\beta F Q(\vec{q})} \omega(F) F }{\int_{F_{\min}}^{F_{\max}} dF\,e^{\beta F Q(\vec{q})} \omega(F) }. 
\end{split}
\label{eqn:fbar}
\end{equation}
At any point in the simulation we can compute $\bar{F}(Q)$ as an additional force to apply to our system and perform the LD scheme in \eqref{eqn:LDq}--\eqref{eqn:LDp}. 
Note that we are free to choose the function $\omega(F)$, but its form will effect the efficacy of the sampling and statistical errors. 
Later we will implement a scheme to learn an efficient $\omega(F)$ on the fly.
From these simulations, it is possible to recover the average of any observable $A$ as if we had performed the simulation with a particular fixed applied force and taken the average over that fixed-force ensemble density:
\begin{equation}
    \rho_F(\vec{q}) = Z^{-1}_q(F) e^{-\beta U(\vec{q}) 
    + \beta F Q(\vec{q})} .
\end{equation}
We can see this by manipulating the equation for $\langle A \rangle_F$ in the following way, to introduce an average over $\bar{\rho}(\vec{q})$ rather than over $\rho_F(\vec{q})$:

\begin{align}
\begin{split}
    \langle A \rangle_F &= \int d\vec{q} \, A(\vec{q}) \rho(\vec{q},F)  \\
                        &\equiv \int d\vec{q}\, A(\vec{q}) \bar{\rho}(\vec{q}) W_F(\vec{q})\\
                        &=\lim_{T\rightarrow{}\infty} \frac{1}{T} \int_0^T dt\, A(\vec{q}(t)) W_F(\vec{q}(t))
                        \end{split}
                        \label{eqn:ergo1}
\end{align}
where in \eqref{eqn:ergo1} we have used the property of ergodicity to transform an ensemble average to a time average over the simulation, and $W_F(\vec{q})= \rho_F(\vec{q})/\bar{\rho}(\vec{q})$ is the observable weight that we will use to recover the correct average of observable $A$ from our simulation. A direct calculation as in Ref.~\citenum{martinsson2019simulated} shows that  $W_F(\vec{q})$ can be expressed in terms of $\omega(F)$ and the as yet unknown partition functions $Z_q(F)$: 
\begin{align}
    &W_F(\vec{q}) = \left [ \frac{\rho_F(\vec{q})}{\bar{\rho}(\vec{q})} \right ]  \\
           & = \frac{ Z^{-1}_q(F) e^{-\beta U(\vec{q}) + \beta F Q} [\int dF' Z_q(F') \omega(F')]}{\int dF' \omega(F') e^{-\beta U(\vec{q}) + \beta F' Q(\vec{q})} }\\
           & = \frac{ Z^{-1}_q(F) \int dF' Z_q(F') \omega(F')}{\int dF' \omega(F') e^{\beta (F'-F) Q(\vec{q})} }
           \label{eqn:oweights}.
\end{align}
Here, we have suppressed the integration range over force for compactness.

Following Ref.~\citenum{martinsson2019simulated}, given a set of weights $\omega(F)$, we can find an expression for $Z_q(F)$ up to a constant factor: 
\begin{align}
    Z_q(F) &= \int d\vec{q}\, e^{-\beta U(\vec{q}) + \beta F Q}\nonumber \\
           &= \int d\vec{q}\, e^{-\beta U(\vec{q}) + \beta F Q} \frac{\bar{\rho}(\vec{q})}{\bar{\rho}(\vec{q})} \nonumber \\
           &= \int d\vec{q}\, \bar{\rho}(\vec{q}) \frac{Ce^{\beta F Q(\vec{q})}}{\int d F' \omega(F')e^{\beta F' Q(\vec{q})}} \\
           &= \lim_{T\to\infty} \frac1T\int_0^T dt  \frac{C e^{\beta F Q(\vec{q}(t))}}{\int d F' \omega(F')e^{\beta F' Q(\vec{q}(t))}}
           \label{eqn:ergo2},
\end{align}
where \eqref{eqn:ergo2} again follows from ergodicity and $C=\int d\vec{q}\, \bar{\rho}(\vec{q})$---since only ratios like $Z_q(F)/Z_q(F')$ matter, this constant $C$ is irrelevant. 

Being able to estimate $Z_q(F)$ from a simulation trajectory gives us a scheme for choosing the $\omega(F)$.
If we want to be able to compute the average $\langle A \rangle_F$ for any $F$ in our desired force range, then  we can assert that an efficient sampling scheme will have all forces sampled with equal probability.
This happens when $\omega(F)\propto Z^{-1}_q(F)$, in which case the PDF of $F$ is given by
\begin{equation}
    P(F) = \frac{Z_q(F) \omega(F)}{\int dF\, Z_q(F) \omega(F)} = \frac{1}{F_{\max}-F_{\min}},
\end{equation}
for $F\in[F_{\min},F_{\max}]$ and $P(F)=0$ otherwise.
We can construct an adaptive scheme to simultaneously learn the weights and estimate the partition functions as in Ref.~\citenum{martinsson2019simulated}, the details of which are given below.

\subsection{Effective Potential}
In the limit where this uniform sampling is achieved, then configurations of $\vec{q}$ will occur with probability $P(\vec{q})\propto \bar{\rho}(\vec{q})$. 
From this probability density function, we can define an effective potential energy that the system samples up to an additive constant as $U_{\rm eff}(\vec{q})=-k_B T \log(P(\vec{q}))$, or 
\begin{equation}
    e^{-\beta U_{\text{eff}}(\vec{q})} \equiv \int_{F_{\min}}^{F_{\max}} d F' \, \omega(F') e^{-\beta U(\vec{q})+\beta F' Q(\vec{q})}.\label{eqn:Ueff}
\end{equation}
This expression is valid for any choice of $\omega(F)$, including $\omega(F) = Z^{-1}_q(F)$;
it can be evaluated using numerical integration for any test potential, and hence we can use it as a reference to predict the expected behavior of our sampling method for those cases. 

\subsection{Algorithm for learning weights}
Above, we discuss that in order to sample all forces with equal probability, the form of the weights are such that $\omega(F) \propto Z^{-1}_q(F)$. 
Here, we sketch the algorithm and implementation details used to learn the weights on-the-fly during sampling, following the same scheme as Ref.~\citenum{martinsson2019simulated} as implemented in the MIST package \cite{bethune2019mist}.

A good numerical scheme for adapting the weights and performing integrals of the form Eq.~\eqref{eqn:fbar} and \eqref{eqn:ergo2} is to learn the weights at a fixed set of $M$ ``node'' points ($f_i$) placed at the roots of a Legendre polynomial between $F_{{\min}}$ and $F_{{\max}}$, and perform the integrals by Gauss-Legendre quadrature; for this we use the implementation of John Burkardt.\footnote{people.sc.fsu.edu/~jburkardt/c\_src/legendre\_rule\_fast/legendre\_rule\_fast.html} Here, each $f_i$ has a corresponding weight $B_i$ such that for a function $g$, $\sum_{i=1}^M g(f_i) B_i \approx
 \int_{F_{\min}}^{F_{\max}} g(F) dF$.

Having chosen an initial distribution of weights, we can begin computing a running average of $Z_q(f_i)$ up to sample number $n$,
\begin{equation}
    z_{i,n} = \frac{1}{n} \frac{e^{\beta f_i Q(\vec{q}_n)}}{\sum_{j=1}^{M} B_j \omega_{j,n} e^{\beta f_i Q(\vec{q}_n)}} + z_{i,n-1} \frac{n-1}{n}.
\end{equation}
We then update the weights at the discrete forces $\omega_i$ by a scheme such that $\omega_i$ converges towards $\omega_i\propto z_i^{-1}$,
\begin{equation}
    \omega_{i,\diamondsuit} = \omega_{i,n} (1-h) + \frac{h}{z_{i,n}},
\end{equation}
and then re-normalize the weights,
\begin{equation}
    \omega_{i,n+1} = \frac{\omega_{i,\diamondsuit}}{\sum_{j=1}^{M} B_j \omega_{j,\diamondsuit}}.
\end{equation}
Here, $h=dt/\tau$, where $\tau$ is a timescale parameter that controls how quickly the weights are adjusted.

We then compute the current average force using the discretized weights by:
\begin{equation}
    \bar{F}(Q) = \frac{\sum_{j=1}^{M} B_j f_j \omega_j e^{\beta f_j Q(\vec{q})}}{\sum_{j=1}^{M} B_j \omega_j e^{\beta f_j Q(\vec{q})}},
\end{equation}
and the observable weights as
\begin{equation}
    W_{f_i}(\vec{q}) = \frac{f_M-f_1}{z_i \sum_{j=1}^{M} B_j \omega_j e^{\beta (f_j-f_i) Q(\vec{q})}}.
\end{equation}

\section{Results}

\begin{figure*}[ht]
\includegraphics[width=\textwidth]{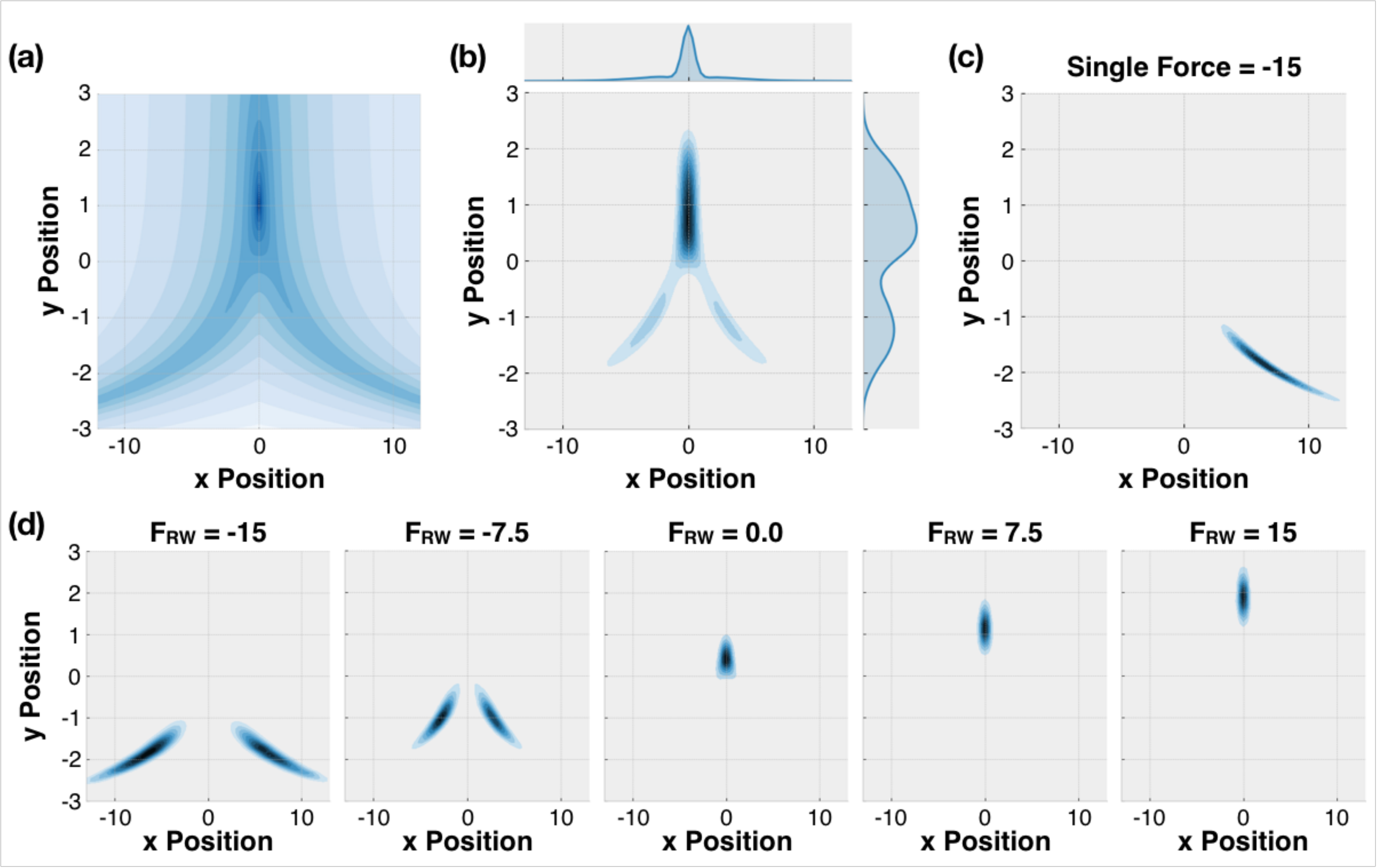}
\caption{(a) Contour plot of potential (\cref{eqn:vpotential}) used for Langevin dynamics (color is plotted on log scale). (b) Total probability density function sampled with FISST using a force range of [-15:15]. Marginal densities for x and y  are plotted on top and right axes respectively. (c) Failure of running a simulation at a single force of -15, where the position is stuck in the right minimum for the entirety of the simulation. (c) Reweighted FISST trajectory to forces of -15, -7.5, 0, 7.5, and 15.}
\label{fig:vpot}
\end{figure*}

In this study, we demonstrate the utility of FISST as a computationally efficient method to simulate force dependent dynamics for a range of systems.
In addition, we show that FISST samples an averaged potential energy surface from the entire range of forces, and discuss how this feature allows FISST to be used as an enhanced sampling method. 
We present results for three different systems of varying complexity: a 2D, V-shaped analytical potential for proof-of-concept, a chain of 12 beads, where interbead interactions are chosen such that the global minimum configuration is degenerate between a left and right handed helix, and deca-alanine in explicit water, which shows that FISST can immediately be applied to atomistic biological systems. 
Together, these systems will be used to illustrate what sampling data is accessible with FISST, identify the accuracy of FISST results, and analyze the performance of FISST relative to alternative methods.

\subsection{Analytical Potential}
We first consider the situation of a particle undergoing LD on an analytical potential. 
Using an analytical potential allows us to easily calculate the exact, bias-dependent potential to compare with sampled data and determine the accuracy of FISST for each applied force.
In addition, we can numerically calculate the effective potential (via \eqref{eqn:Ueff}) that is sampled for a given FISST force range.
The potential we have crafted is a V-shaped analytical potential, parameterized such that there is a single minimum when pulling in the $y$ direction, and two minima separated by a barrier when pulling in the $-y$ direction (\cref{eqn:vpotential}).
This minimalist model is chosen to represent the case of a peptide with multiple possible conformations that collapse to a single extended state when pulled upon, as considered in sections below.

\begin{equation}
    \beta U(x,y)=-8\ln\left[e^{-\frac12(x-e^{-y})^2}+e^{-\frac12(x+e^{-y})^2}\right] + \tfrac12y^2
    \label{eqn:vpotential}
\end{equation}

The total sampling with FISST is clearly different from what would be expected of a standard simulation on this potential, where the sampling of the miniumim is elongated relative to the potential and two minima not present in the unbiased potential are formed in the negative $y$ direction (\cref{fig:vpot}a and \cref{fig:vpot}b).
This deviation arises because FISST samples an effective potential defined by \cref{eqn:Ueff}, which is an average over the whole force range. 
Qualitatively, the elongation emerges because applying a force essentially tilts the potential energy surface in the direction of the pulling coordinate.\cite{bustamante2004mechanical} 
In the case of the potential shown in \cref{fig:vpot}, the effective potential has a contribution from each force that tilts the potential in both the $y$ and $-y$ directions. 
Using the observable weights defined in \cref{eqn:oweights}, we reweight the FISST trajectory to forces, F = -15, -7.5, 0, 7.5, and 15, and plot each probability density in \cref{fig:vpot}d.
The effect of tilting in both directions is evident from each reweighted density, where large negative forces sample in $-y$ and large positive forces sample in $y$.

This potential was chosen such that a strong force in the $-y$ direction would have two degenerate minima, and at the largest negative force considered, $F=-15$, FISST samples each of these minima approximately equally. 
However in trajectories with a single applied force, a force of $F=-15$ leads to only one of these wells being sampled (\cref{fig:vpot}c).
The large force in the $-y$ direction deepens each of these two minima and the particle is unable to escape the first minimum that it samples.
This clearly shows a significant advantage of FISST, where sampling over an effective potential flattens force-specific energy barriers across the potential energy surface without limiting the ability to reweight the trajectory to a specific force, thereby improving sampling at each force.

While it is clear for this example that reweighting a FISST trajectory to different forces across the force range qualitatively reproduces the expected potential energy surface from individual simulations under a constant force, we quantify the error of each reweighted density by calculating the Jensen-Shannon distances\cite{schindelin2003jsd} against the exact probability density (\cref{fig:vpoterror}, for $N_f=20$).
These results show that each reweighted density from FISST is within error of a single force calculation run over the same simulation timescale.
If the length of the single force trajectories are reduced such that the total work for collecting the FISST and single force data sets is the same, we see that the error from FISST at any given force is far less than that from those single force trajectories.

\begin{figure*}[t]
\includegraphics[width=\textwidth]{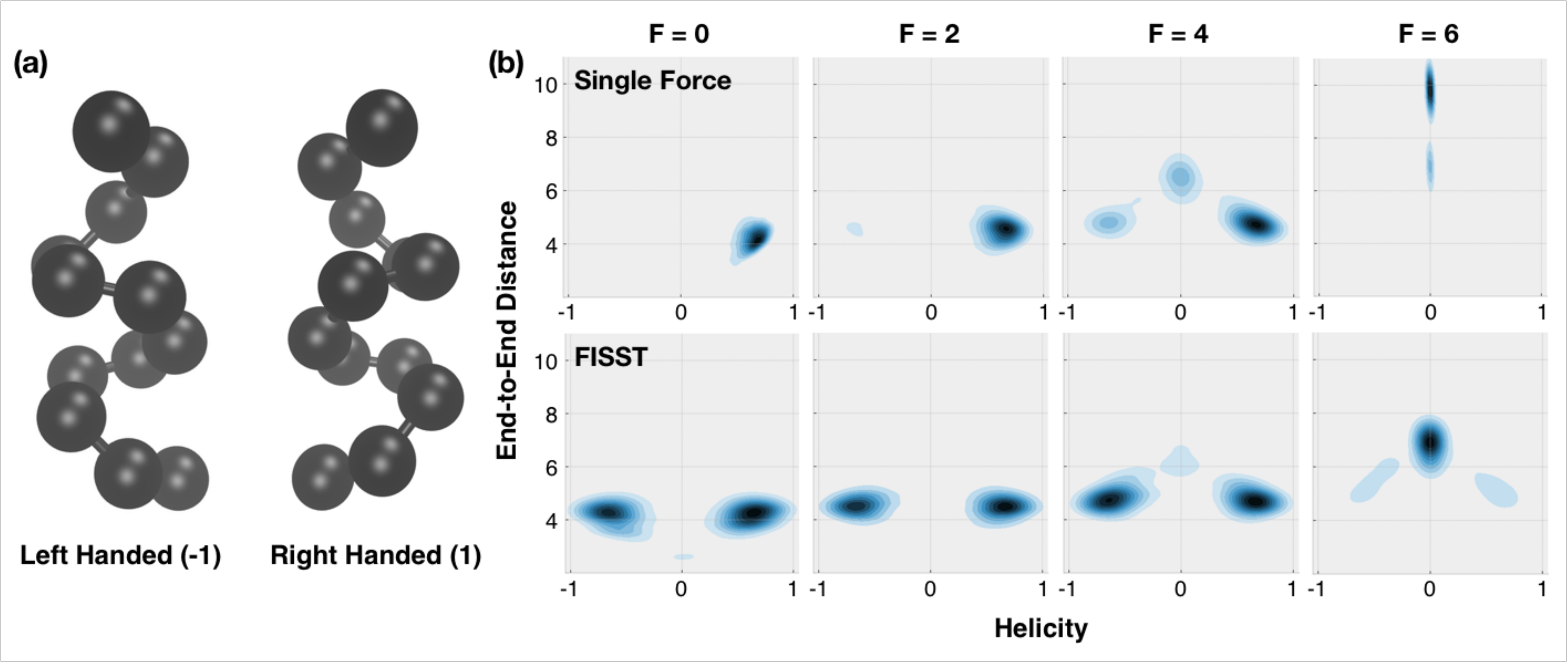}
\caption{(a) Left and right handed helices used as reference structures to quantify helicity. (b) Potential of mean force mapped onto end-to-end distance and helicity for of Single force (top) and FISST (bottom) at F = 0, 2, 4, and 6.}
\label{fig:helix}
\end{figure*}

\subsection{Beaded Helix}
We now consider a toy model of an $\alpha-$helix, designed to be a higher-dimensional analog of the V-shaped potential considered above. 
This helix is composed of a 12 atom chain of beads, where each bead is connected by a harmonic spring of length 1 and spring constant $k=100 k_B T$/distance$^2$.
Additionally, we apply a Lennard-Jones interaction with a strength of $\epsilon=6.0$ $k_B T$ between the ith and i+4th beads, and a purely-repulsive WCA potential \cite{weeks1971role} with $\epsilon=3 k_B T$ between all other bead pairs.
This setup leads to a model with two degenerate ground state conformations of a left or right handed helix (\cref{fig:helix}a), analogous to the analytical potential discussed above.
In these simulations, we apply a pulling force to the terminal atoms of the helix and plot the 2D probability density of end-to-end distances and helicity for the system.
The helicity is determined from the minimum RMSD of the frame against a left and right handed helix, where 1 (-1) is a right (left) handed configuration, and 0 represents extended structures with RMSDs greater than 1.0 for both references.

The sampling of the helix with no applied force shows the same behavior that was seen in the Fig.~\ref{fig:vpot}(c) for the analytical potential, where sampling of only one folded minimum is observed.
This behavior is expected because the interaction energy between beads is chosen to be high relative to the temperature, causing the system to get stuck in the initial right-handed configuration and at no point over the course of the trajectory does the helix unravel enough to switch handedness. 
A pulling force to elongate the helix is required to aid the system in unfolding and allow a change in handedness of the helix. 
Unfortunately, if a relatively high force is used ($F=6$) with a conventional method, only extended conformations are observed (\cref{fig:helix}b, top panel). 
Applying a single force can help get the system unstuck from the initial configuration, however it doesn't evenly sample each folded configuration, because converging sampling of these now-low-probability states is difficult.
The effective potential from FISST provides an alternative that allows proper sampling of the unbiased PES of this helical system, because it flattens the potential at extremes of the force range. 
High pulling forces are present to pull the helix out of its initial right handed configuration and low forces effectively act as restoring forces to allow the helix to refold in either conformation with equal probability. 
Using FISST we observe a relatively balanced population of left and right handed helices when reconstructing the zero force PES of the system (\cref{fig:helix}b bottom), matching what is expected from these degenerate configurations. 
Importantly, we note that these examples illustrate a hallmark of enhanced sampling methods, where sampling along the helical coordinate is enhanced from a pulling force applied to the ends of the helix. 

\subsection{Alanine-10}

\begin{figure}[ht]
\includegraphics[width=\columnwidth]{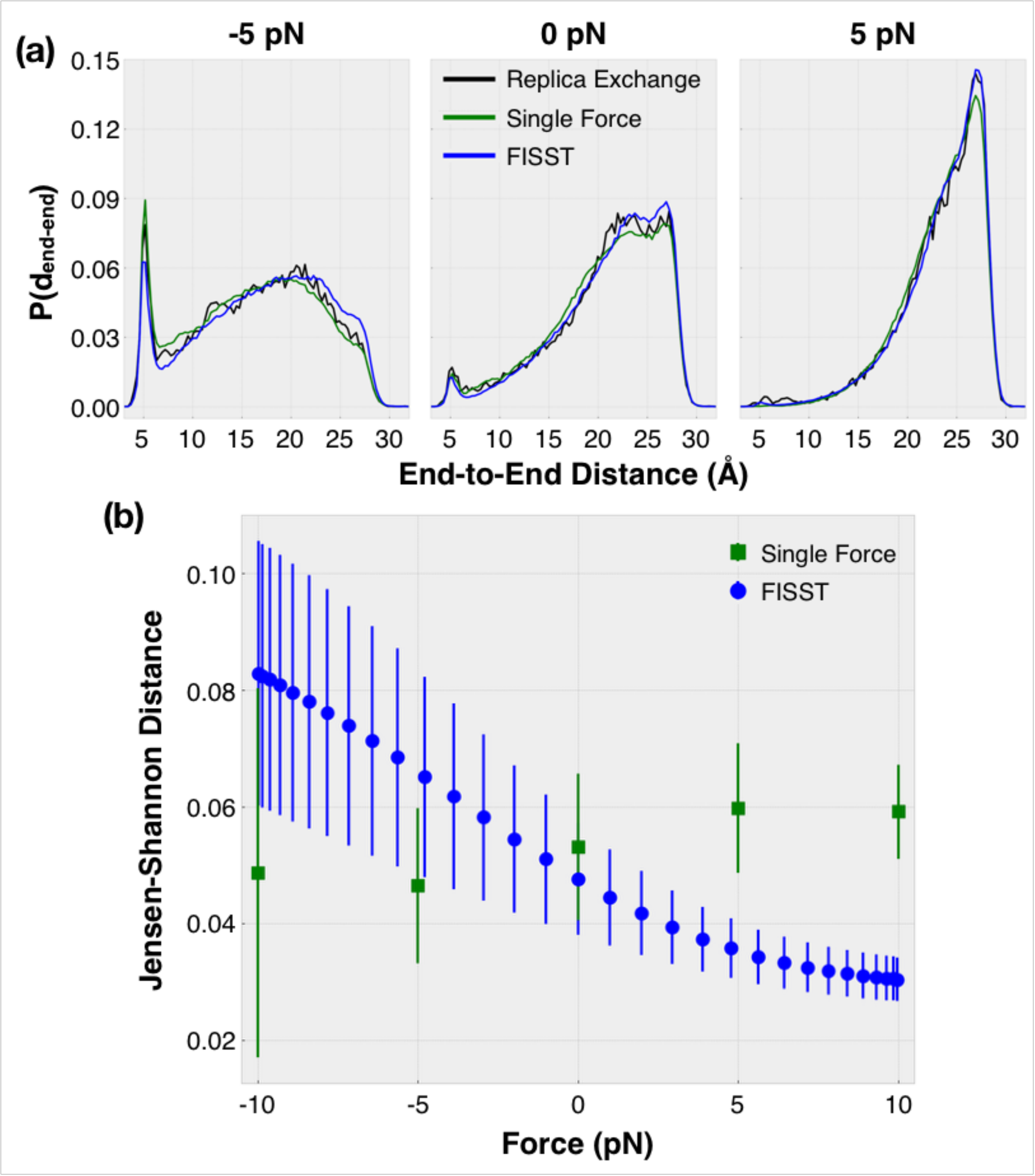}
\caption{Probability density function of end-to-end distances at $F = -$5, 0, and 5 pN calculated from replica exchange, single force, and FISST simulations. (b) Jensen-Shannon distances as a function of force for single force and FISST simulations. The reference density is determined from 5 replica exchange simulations at -10, -5, 0, 5, and 10 pN and interpolated to desired force using EMUS.\cite{thiede2016emus}}
\label{fig:ala10error}
\end{figure}

Up to this point we have considered toy models that illustrate how FISST can be used to both simultaneously sample dynamics at a range of forces through reweighting the trajectory as well as be used as an enhanced sampling method to 'unstick' the system from certain low energy conformations.
The final system we consider is the deca-alanine peptide explicitly solvated in water, which provides a complex, biologically relevant test system for FISST.

To establish benchmark sampling data for alanine-10, we perform temperature replica exchange simulations at forces of -10, -5, 0, 5, and 10 pN, using 40 replicas between 300K and 400K (see \cref{sec:simulations}).
The probability density of end-to-end distances for each method at -5, 0, and 5 pN are shown in \cref{fig:ala10error}a, where all simulations were run for 160 ns.
In \cref{fig:ala10error}a, a separate single force simulation was performed for each force and the FISST data acquired by reweighting with the observable weights determined on-the-fly during a simulation with force range [-10:10] pN.
It is clear that there is a qualitative agreement between the replica exchange benchmarks and both the single force and FISST simulations.

In order to determine the error in the end-to-end distance distribution quantitatively for each method, we employ the Eigenvector method for Umbrella Sampling (EMUS) to interpolate the replica exchange data to other forces for comparison with single force and FISST (\cref{fig:ala10error}b).\cite{thiede2016emus}
This was done by using the replica exchange trajectories at $F = -10$, -5, 0, 5, and 10 pN as input distributions to EMUS and a target force between -10 and 10 pN was provided for EMUS to predict the equilibrium sampling at that force.
This predicted density was then compared to end-to-end densities generated by both single force and FISST simulations at the target force (each point is average over 5 replicates, error bars are $\pm$ one standard deviation of the replicates).

In both FISST and single force simulations, the large standard deviations at negative forces reflect the fact that  compressive forces ($F < 0$) are harder to sample accurately. 
This is due to a larger variety of transient conformations that can be form when the peptide is being compressed, for example bending the chain in ways that cause a uncommon geometries.
However, similar to the results from the V-shaped analytical potential, the average error in single force calculations is relatively constant across the forces considered.
The error in the reweighted FISST calculations decreases from about 0.08 to 0.03, and crosses the average error from running individual single force simulations (\cref{fig:ala10error}b). 
Each single force calculation had the same duration as each FISST simulations, hence the same computational time for single force would only give one data point in this range, whereas we get all points simultaneously with FISST.
Despite this large decrease in total computational cost to collect the FISST data shown, the calculated end-to-end density with FISST have approximately the same total error and standard deviation from multiple replicates as single force calculations.

\begin{figure}[ht]
\includegraphics[width=\columnwidth]{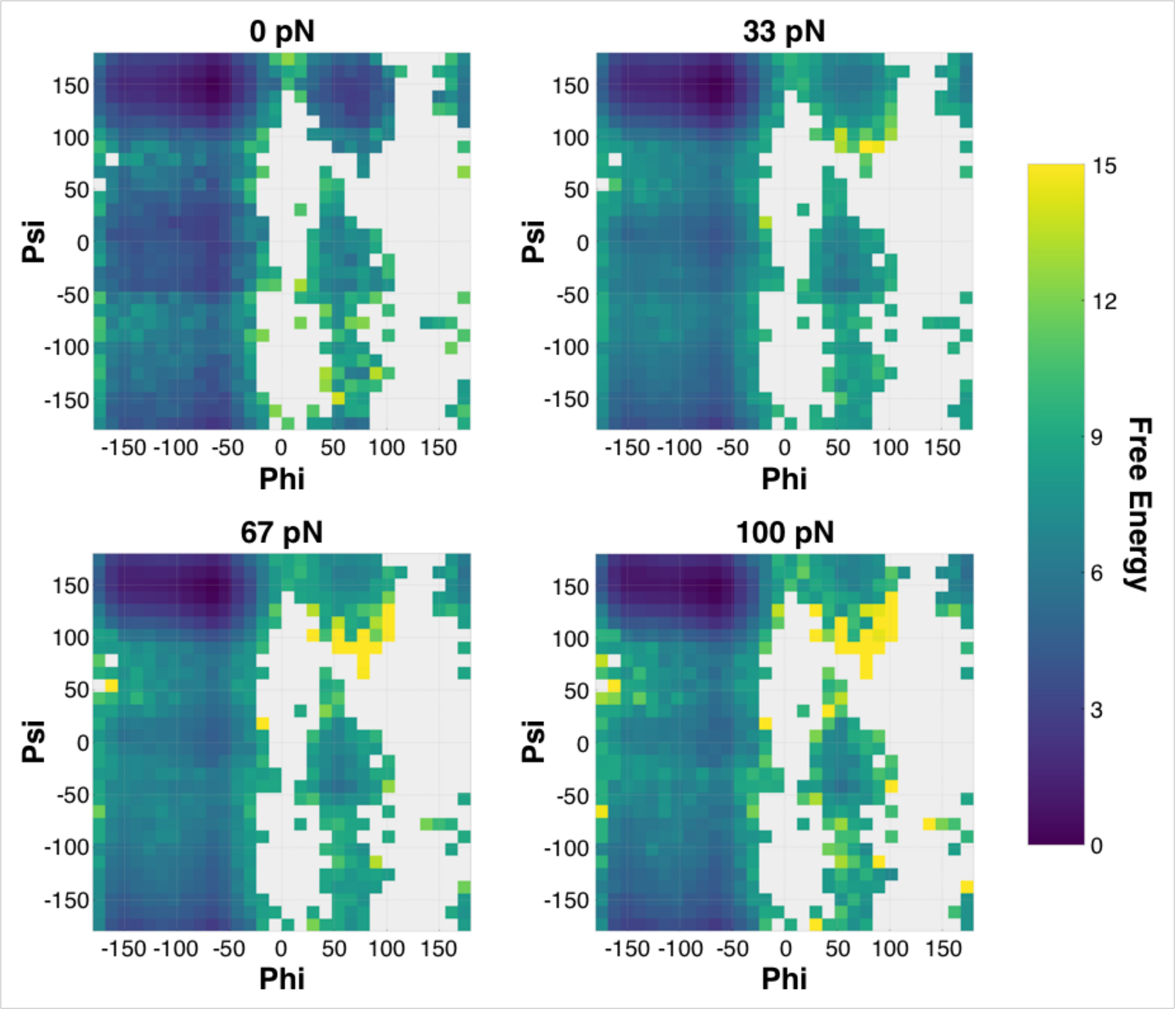}
\caption{Ramachandran plots of alanine-10 peptide at 0, 33, 67, and 100 pN applied force. Each plot was generated from a single, 160 ns FISST simulation using a force range of [0:100] pN.}
\label{fig:rama}
\end{figure}

Lastly, we wish to illustrate that the observable weights from FISST can allow one to reconstruct the probability densities of other observables, not just the one that was biased. 
In \cref{fig:rama}, we show the effect of pulling on the end-to-end distance on the dihedral angle densities of the alanine-10 peptide. 
Small alanine peptides are known to prefer the polyproline II (PPII) helix, a left handed helix that is be present in many folded, unfolded, and amorphous biomolecules.\cite{makarov2013ppiihelix}
\cref{fig:rama} shows the Ramachandran plots of alanine-10 at 0, 33, 67, and 100 pN calculated using FISST with a force range of [0:100] pN, where the lowest energy configuration is referenced to zero.
Analogous Ramachandran plots calculated with single force and replica exchange simulations are shown in \cref{fig:sframa} and \cref{fig:rerama}, and qualitatively agree with the FISST data. 
The global minimum conformation under each applied force is the polyproline II (PPII) region at (-75, 150) degrees, consistent with the literature of short alanine peptides.\cite{Kallenbach2002ppii}
Additionally, there is a local minimum in the $\alpha$-helical region of the plot, consistent with previous work on alanine peptides, where the $\alpha$-helix is known to be a stable conformation.\cite{schwalbe2007alanine19}
For all methods used, we observe that as force is increased, the helical basin decreases in population relative to the less compact PPII structure, as shown by previous theoretical and computational work \cite{stirnemann2013elasticity,guerin2018conformational}.

\section{Discussion and conclusions}

In this work, we present the FISST method for performing simulated tempering in force in the infinite switch limit.
In FISST, observable weights are computed during the simulation which can be used to compute averages of any structural quantity at any force within the simulated force range.
We benchmark this new method on a variety of model systems with varying complexity to evaluate the performance of FISST, including a simple analytical potential, a toy helix, and deca-alanine in water.
For each system we showed that a FISST simulation was able to quantitatively reproduce the quality of sampling at each force across the force range at lower total computational cost, and in some cases accelerated sampling over barriers that could not be crossed in a standard MD simulation.
This efficiency makes FISST a promising method for studying the response of larger and more complex biomolecular systems to small applied forces. 

A key point of entry for using FISST it to choose a system specific force range that is relevant to the problem being studied.
Through testing FISST, we have found that the quality of sampling can depend on the choice of force range, however simple intuition about the system is usually sufficient to overcome these differences. 
For example, we have observed that enhanced sampling at $F=0$ can be aided using force ranges that extend into the negative (compression) region to include restoring forces.
FISST causes the molecule to always feel some effect from every force in the force range.
This can lead to some problems with very high force, especially before the weights have converged at the beginning of the simulation, where the large forces can contribute more that expected.
However, after only a couple nanoseconds, the weights converge and remain stable for the remainder of the simulation (\cref{fig:weights}). 
While running long enough simulations will eliminate this effect, another simple option for studying very wide force ranges is to run multiple FISST simulations over using smaller force ranges that span the target force difference. 
The current implementation also includes the ability to select different initial weight distributions, such that large forces have small weights at the beginning, however that was not necessary for any of the examples in this current work. 

As currently implemented and described above, FISST can only be applied to a single collective variable.
The formulation herein can trivially be extended to higher dimensions, however as with many similar histogram methods, the need to learn the weights over a discrete set of points means that it is not likely to perform well for more than two dimensions. 
Rather than going to higher dimensions in FISST, we believe the most promising strategy is to apply FISST to study mechanical forces along a CV of interest, and combine that simulation with other methods that will accelerate the sampling of conformations along other degrees of freedom. 
With the weights fixed, FISST is a fully equilibrium sampling method, hence any other equilibrium method (such as various forms of tempering, umbrella sampling, metadynamics, variationally enhanced sampling, etc. \cite{marinari1992simulated,fukunishi2002hamiltonian, wang2011replica, torrie1977nonphysical, laio2008metadynamics, barducci2008well, valsson2014variational}) can be used on top of the learned FISST potential. 
We are currently exploring which of these other enhanced sampling methods can be rigorously combined during the FISST simulation to accelerate the convergence of the sampling during the time when the weights are being learned. 

Lastly, we note that, although we have targeted the problem of understanding the effect of mechanical forces along a collective variable, FISST could be used to accelerate the sampling across any range of linear coupling terms in a Hamiltonian. 
Because FISST is implemented in PLUMED, it can be immediately applied to any CV to enhance sampling over a range of couplings for that CV, whether or not it corresponds to a physical force. 
As one example, FISST could be applied along with the DIPOLE CV to enhance sampling over a range of electric fields. 
Additionally, recent studies have shown that experimental information can be directly incorporated into MD simulations with minimal bias using extra linear coupling terms in the Hamiltonian \cite{pitera2012use, white2014efficient, hocky2017coarse, cesari2018using, amirkulova2019recent}. 
It will be interesting to explore whether this method, which flattens the probability of seeing a given coupling term, can be connected to understanding the optimal terms determined by those relative-entropy based methods. 

\section{Simulation details}
\label{sec:simulations}
All code for the method is currently being contributed as a module in the PLUMED open source sampling library \cite{Bussi2014plumed2}. The version used for this work is available in the FISST-dev branch in our group's github repository (\url{https://github.com/hocky-research-group/plumed2}). Input files and scripts for repeating the calculations in this work will be deposited in the PLUMED-NEST (\url{https://www.plumed-nest.org/}) \cite{bonomi2019promoting}. 

\subsection{Langevin dynamics on test potentials}

The Langevin dynamics on the analytical v-shaped potential defined in \eqref{eqn:vpotential} were performed using the pesmd module implementation in PLUMED. 
The simulation was initialized at position (0.0, 0.0), with a temperature of 1.0 $k_BT$, a timestep of 0.05, and friction of 1 was used. 
A total of 5,000,000 steps were collected, where every 20th frame of the simulation was used for the analysis. 
The FISST simulations were run with a force range of [-15:15], 21 quadrature points were used to discretize the force range, a uniform initial weight distribution was used, and the weights were updated every 200 steps.

\subsection{Atomistic molecular dynamics on peptides}

All atomistic molecular dynamics simulations were done using GROMACS version 5.1.4.\cite{lindahl2015gromacs, berendsen2005gromacs} 
Alanine-10 was set up in explicit TIP3P water and parameterized using the CHARMM36 all atom forcefield\cite{mackerell2012charmm36}. 
A cubic box with edge length 57.0 \AA was used and periodic boundary conditions were used in all three directions. 
The initial structure was minimized using the steepest descent algorithm for a maximum of 50000 steps. 
The cutoff for short range interactions was chosen to be 1.0 \AA. 
Longer range coloumbic interactions were coupled using the Particle Mesh Ewald method and constraints for hydrogen bonds were computed using the LINCS algorithm.\cite{fraaije1997lincs}
Minimization was first followed by a NVT equilibration using the Berendsen thermostat\cite{haak1984berendsen}, then a NPT equilibration where pressure was maintained at 1.0 atm using a Parinello-Rahman Barostat\cite{rahman1980npt} and Bussi-Parinello thermostat\cite{parinello2007vrescale}.

Constant pressure FISST simulations of deca-alanine with force ranges of [-10:10] pN and [0:100] pN were run to collect the data in \cref{fig:ala10error} and \cref{fig:rama} respectively. 
In each case, 31 quadrature points were used, the weights were initiated with a uniform distribution, and a period of 200 steps was used to update the weights.

Each force dependent replica exchange simulation was performed at 40 different temperatures between 300 and 400 K. 
The production molecular dynamics runs were carried out in the NPT ensemble. 
Each time step was 2.0 fs and the equations of motion were integrated using the verlet algorithm. 
The pulling for each force was implemented using PLUMED's RESTRAINT function, where force was applied on the terminal $\alpha$ carbons.
The total simulation time for each production run was 160 ns.

\begin{acknowledgement}
Simulations for this work were performed on resources provided by NYU IT High Performance Computing. We thank Jonathan Weare for helpful conversations, and Erik Thiede for providing an example script for reweighting using EMUS. We thank Ian Bethune and Anton Martinsson and Benedict Leimkuhler for sharing their ISST implementation in the open-source MIST software package. 

\end{acknowledgement}

\bibliography{fisst}

\clearpage
\onecolumn

\section{Supporting Information}
\setcounter{figure}{0}
\renewcommand{\thefigure}{S\arabic{figure}}
\begin{doublespace}

\subsection{V-Shape Potential}

\begin{figure}[h]
\includegraphics[width=\textwidth]{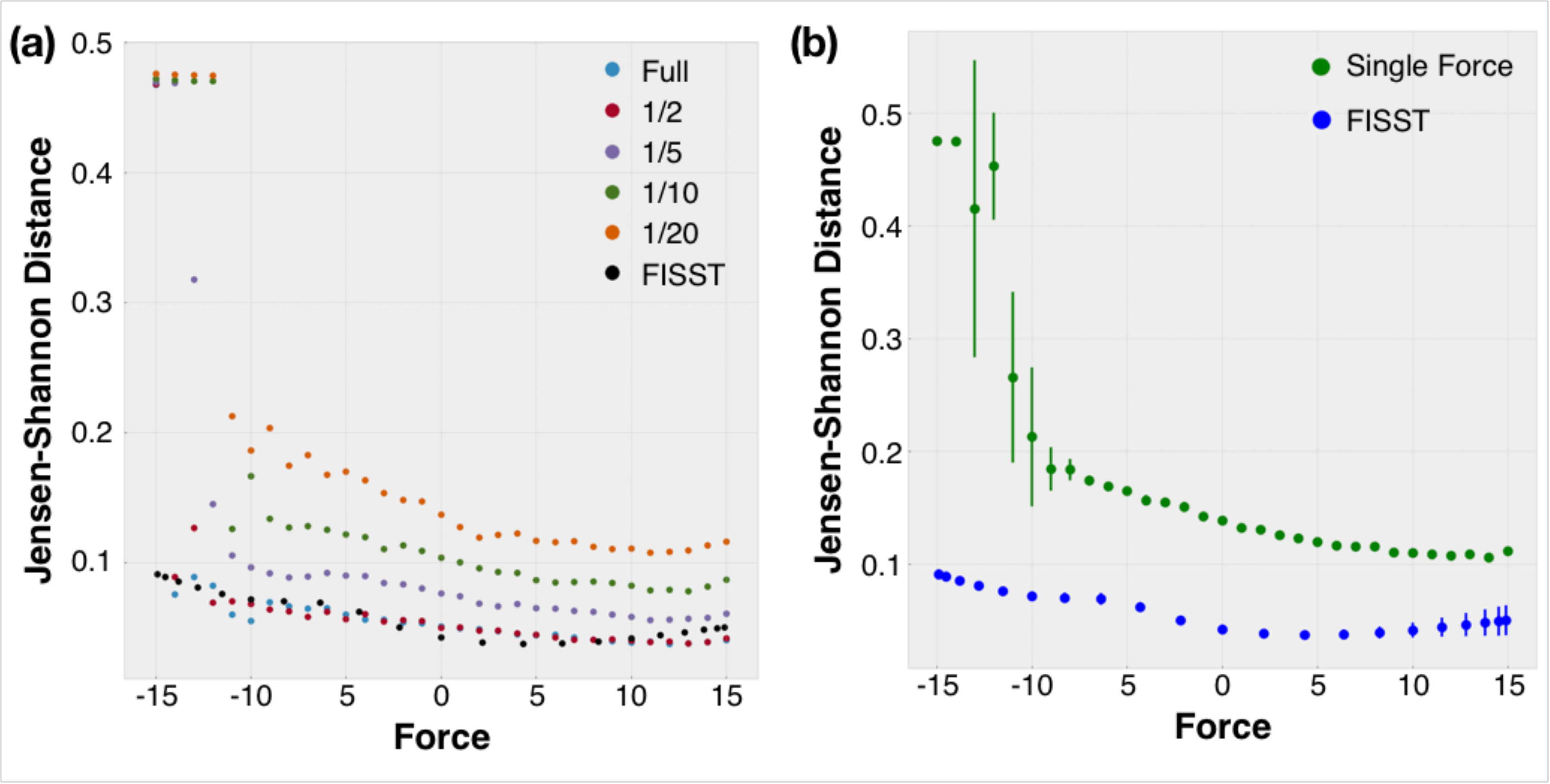}
\caption{(a) Error relative to exact density of FISST [-15:15] and individual single force simulations run with a fraction of the steps. In the 1/20 data set the error is very high compared to FISST despite the number of total steps across all forces being equal to that of the full FISST trajectory. Simulations on the order of the same length of FISST are required at each single force to replicate the same level of accuracy. (b) Average error relative to exact density over 5 replicates for FISST [-15:15] and single force simulations run with 1/20th the number of steps.}
\label{fig:vpoterror}
\end{figure}

The most simple alternative to FISST is doing a separate simulation at each force. 
For the v-shape potential considered in the main text, this brute force approach leads to error that, other than large negative forces, is more or less constant across the force range studied, decreasing slightly at larger positive forces.
In order to evaluate how efficient FISST is over traditional approaches, we ran single force simulations at a fraction of the length of FISST and plotted the error as the trajectory is shortened (\cref{fig:vpoterror}a).
Because we are considering 20 different forces, the $\frac{1}{20}$th dataset (orange points) contains the same total work as the FISST trajectory (black points).
In addition to getting the large negative force regime qualitatively correct, this data shows that FISST is much more efficient than standard methods while not compromising the accuracy of the simulation.

\cref{fig:vpoterror}b plots the average error for a constant amount of work for both FISST and single force simulations averaged over 5 simulations.
Here, the quality of the simulations are constant over multiple independent simulations.
In the single force case, larger deviations between trials show up at negative forces ($<-10$), which is the point at which these simulations begin to fail due to getting stuck in one of the arms of the potential.
In none of the 5 independent trials at $F=-14,-15$ did a single force simulation sample both minima.

\subsection{Alanine-10}

\begin{figure}[ht]
\includegraphics[width=\textwidth]{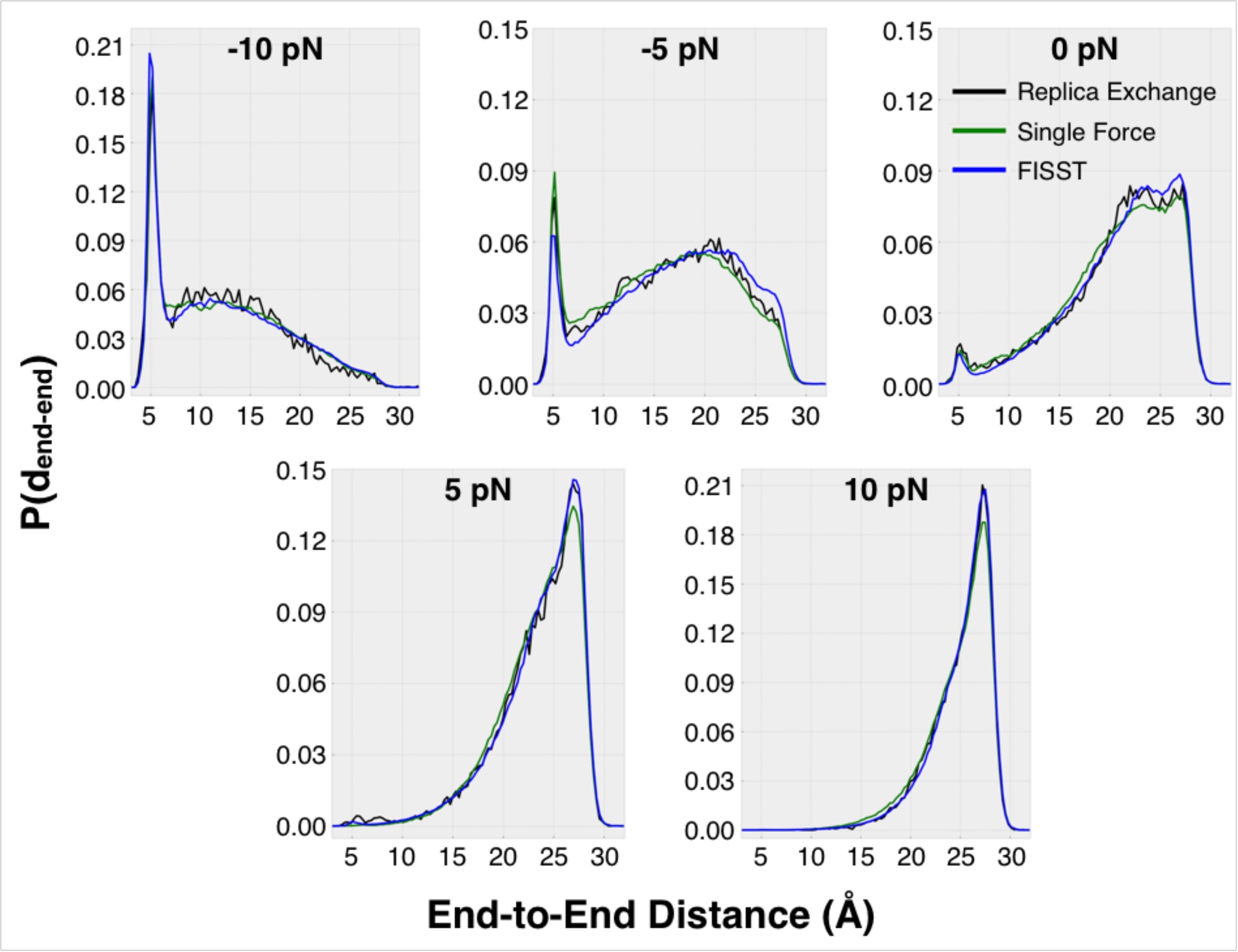}
\caption{Probability density of end-to-end distances at F = -10, -5, 0, 5, and 10 pN calculated from replica exchange, single force, and FISST simulations.}
\label{fig:e2edistall}
\end{figure}

In total, we performed replica exchange simulations at -10, -5, 0, 5, and 10 pN to get a set of benchmark simulations to compare to FISST and single force simulations run over the same force range.
Here, \cref{fig:e2edistall} shows the comparison between each method for each RE simulation run under force.
At each force to which the FISST data is reweighted, we clearly match the end-to-end distance probability density from the replica exchange benchmark simulations. 
Each of the single force simulations are run for the same timescale as the single FISST run, however a separate simulation was required to calculate the probability density at each force. 

\clearpage

\begin{figure}[ht]
\includegraphics[width=\textwidth]{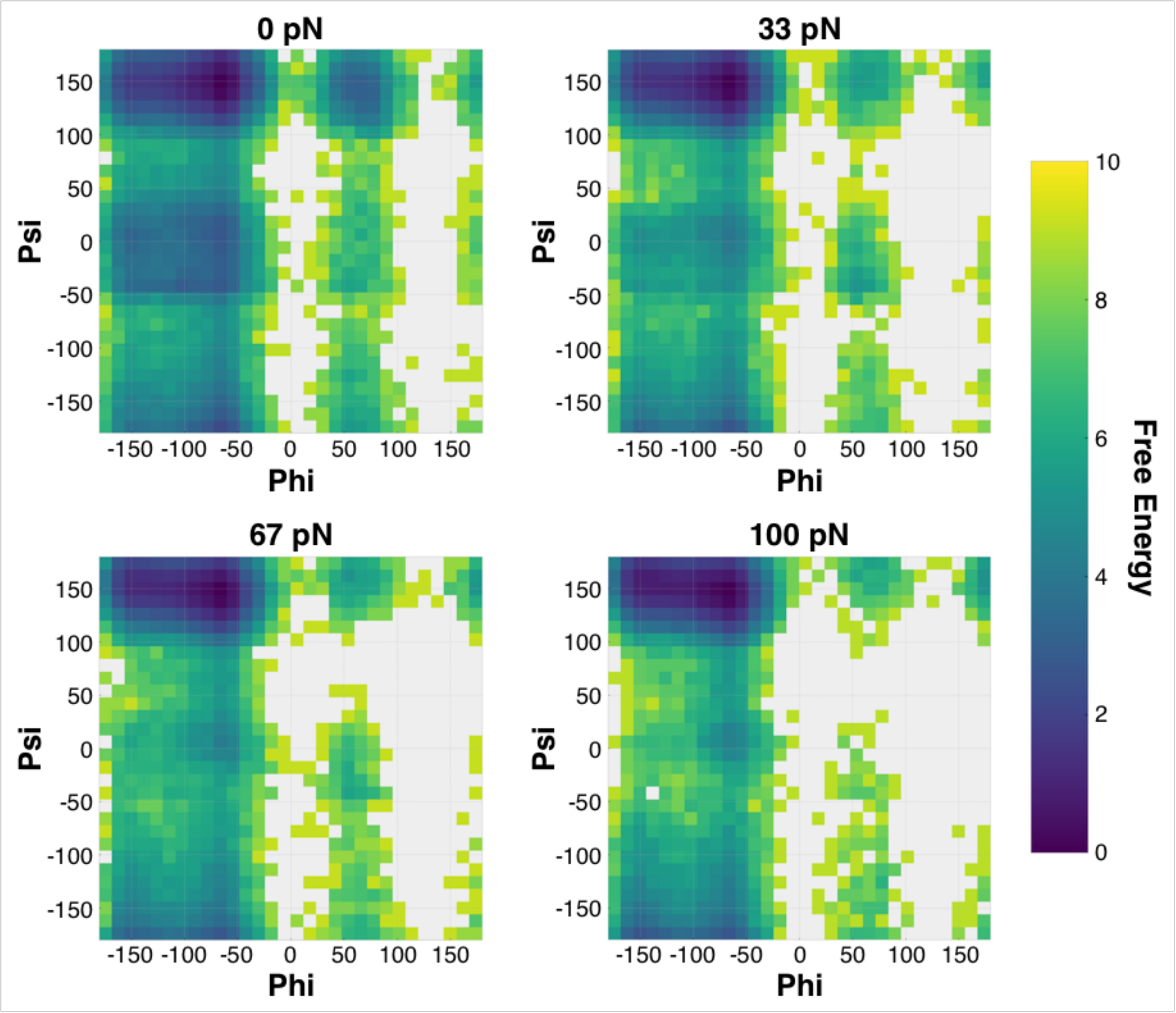}
\caption{Ramachandran plots of alanine-10 peptide at 0, 33, 67, and 100 pN applied force, each simulated at a single applied force.}
\label{fig:sframa}
\end{figure}

\begin{figure}[ht]
\includegraphics[width=\textwidth]{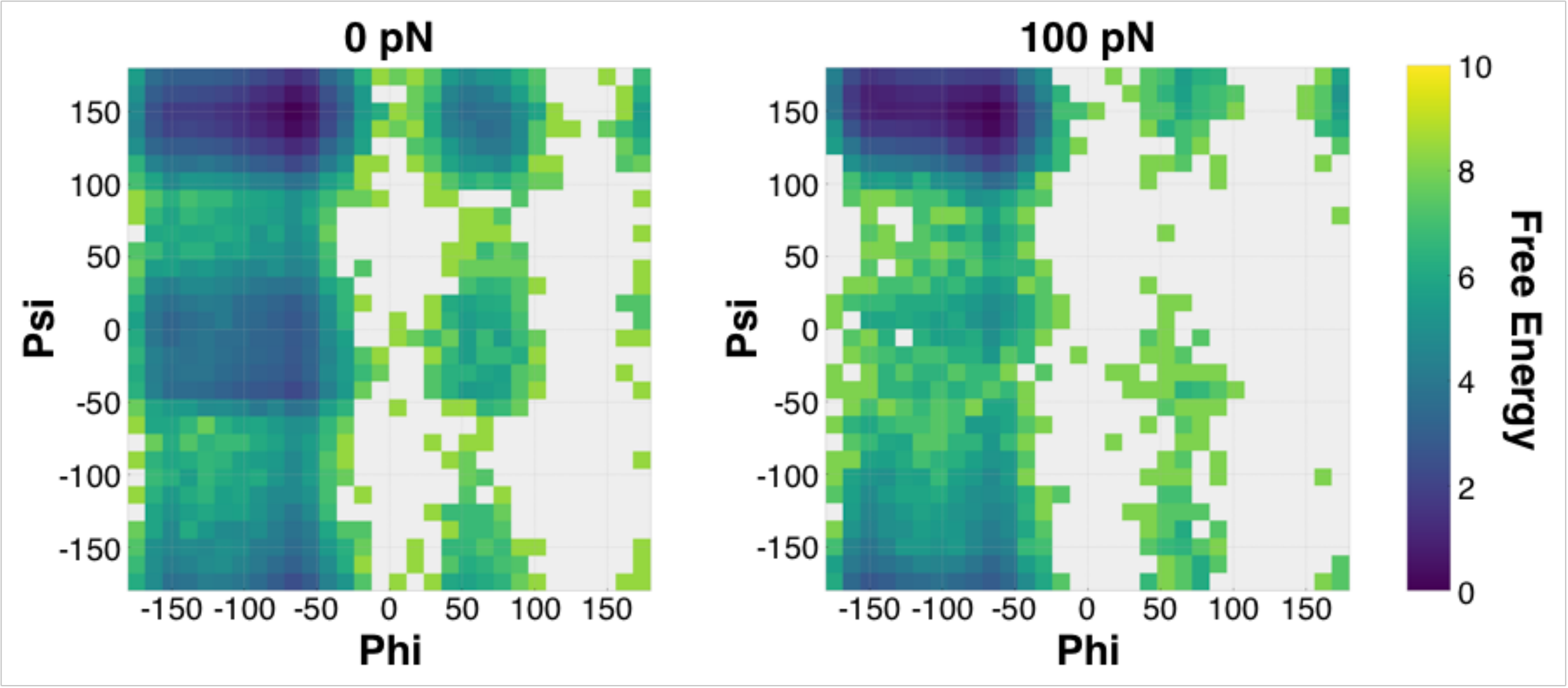}
\caption{Ramachandran plots of alanine-10 peptide at 0 and 100 pN applied force, each calculated from Replica Exchange simulations.}
\label{fig:rerama}
\end{figure}

\clearpage

\begin{figure}[ht]
\includegraphics[width=\textwidth]{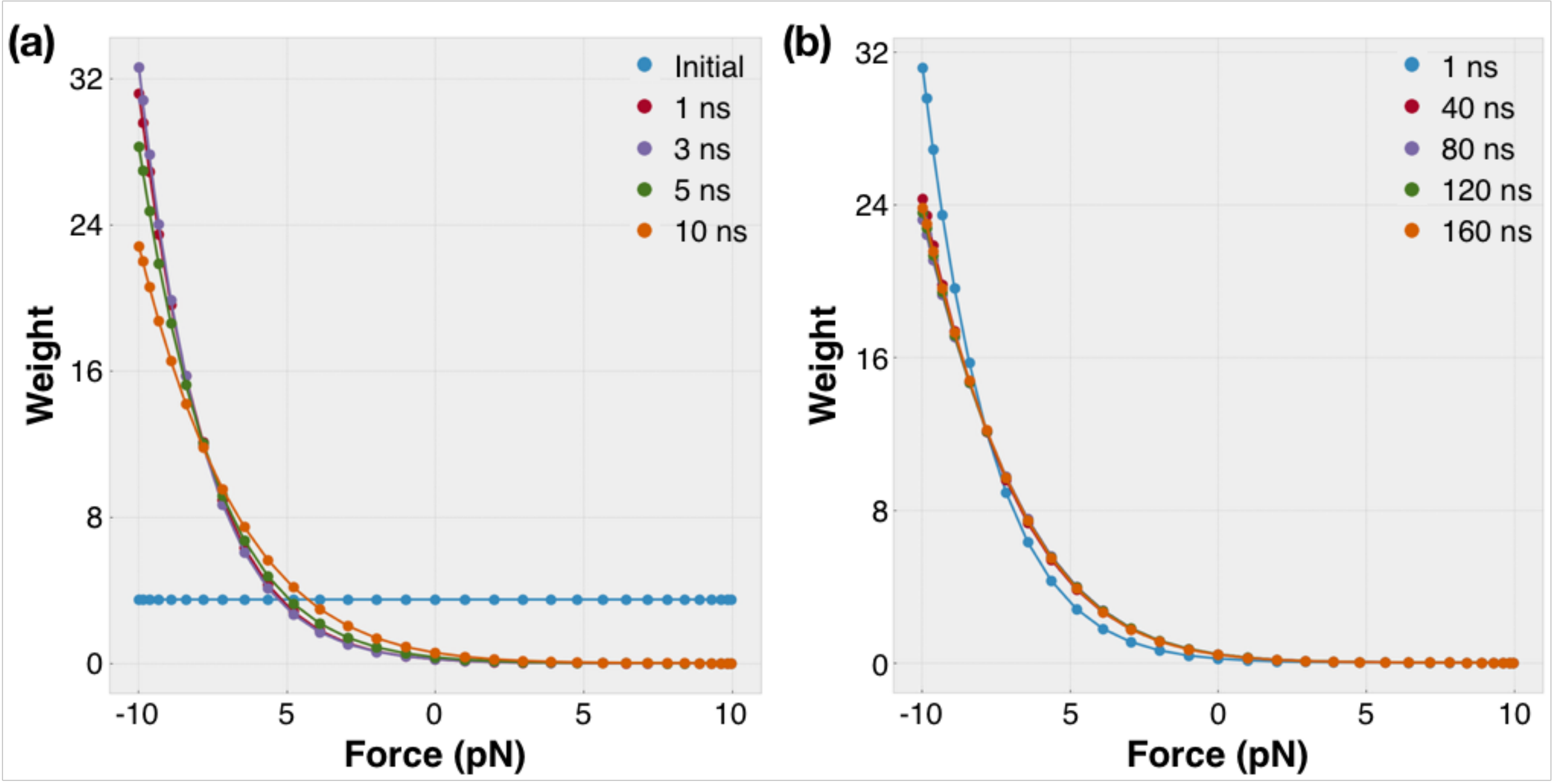}
\caption{Distribution of weights at various stages of a 160 ns deca-alanine simulation. (a) At short times, the weight distribution changes significantly. (b) After 40 ns, the weight distribution is essentially constant for the remainder of the simulation.}
\label{fig:weights}
\end{figure}

In \cref{fig:weights} we plot the distribution of the weights as they are being learned during a 160 ns deca-alanine simulation.
The initial weight distribution is chosen to be uniform and it quickly changes to an approximately exponential distribution at short times.
During the first 10 ns of the simulation, both the amplitude and decay rate of the distribution change relatively quickly.
After this initial learning stage, the weight distribution does not significantly change for the remainder of the simulation. 
\clearpage

\begin{figure}[ht]
\includegraphics[width=\textwidth]{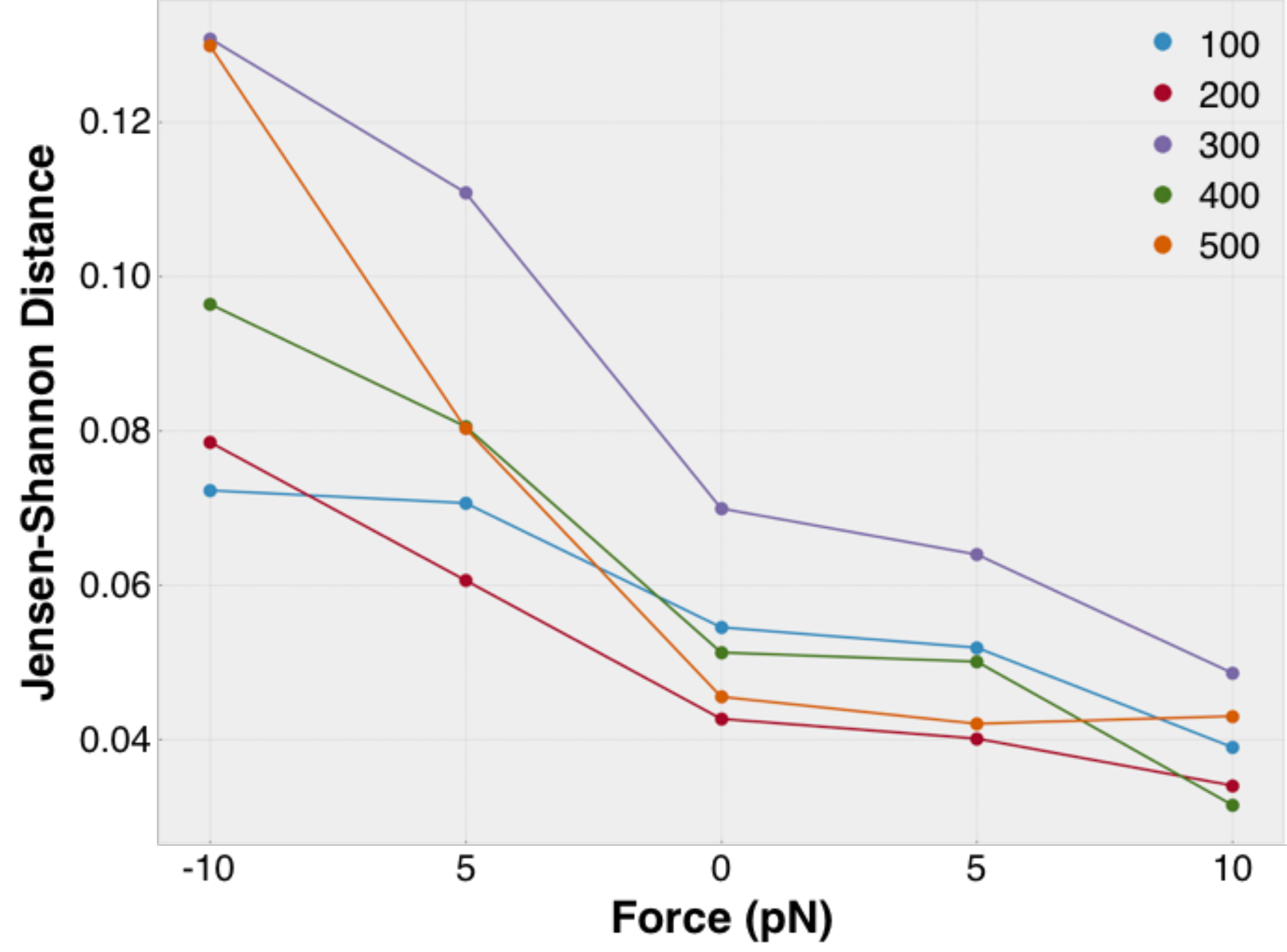}
\caption{Error in FISST for a range of learning rates. A 160 ns FISST simulation was run with the weights updated every 100, 200, 300, 400, and 500 steps. We calculate the end-to-end distance probability densities at each force and calculate the error against results from replica exchange simulations. In the range considered, there is no obvious dependence on learning rate. An update period of 200 was used for all data discussed in the main text.}
\label{fig:period}
\end{figure}

In \cref{fig:period} we plot the force dependent error in FISST simulations calculated with 5 different learning rates.
The chosen learning rate does not significantly impact sampling over the range of learning rates considered.

\end{doublespace}

\end{document}